\documentclass[dvips, 12pt]{article}

\usepackage{amsmath, amssymb, amsthm}
\usepackage[bf, small]{caption}
\usepackage{epic}

\newcommand {\rulefig}[2] {
    \begin{figure}[h]
        \hrule
        \vspace{-5pt}
        #1
        \hrule
        \caption{\textbf{#2}}
    \end{figure}
}

\newcommand {\fig}[2] {
    \begin{figure}[h]
        \setlength{\unitlength}{0.030in}
        #1
        \caption{\textbf{#2}}
    \end{figure}
}

\newcommand {\half} {{1 \over 2}}
\newcommand {\hhalf} {{\textstyle{1 \over 2}}}

\newcommand {\rthalf} {{1 \over \sqrt 2}}

\newcommand {\ket}[1] {\left| #1 \right>}

\newcommand {\longto} {\longrightarrow}

\newtheorem{thm}{Theorem}[section]
\newtheorem{cor}[thm]{Corollary}
\newtheorem{lem}[thm]{Lemma}
\theoremstyle{remark}

\theoremstyle{definition}
\newtheorem{defn}[thm]{Definition}

\newcommand {\bcode}{\addtolength{\jot}{-3pt}
                     \begin{equation*}}
\newcommand {\ecode}{\end{equation*}
                     \addtolength{\jot}{+3pt}}
\newcommand {\brules}{\vspace{4pt}\addtolength{\jot}{+4pt}}
\newcommand {\erules}{\addtolength{\jot}{-4pt}
                      \vspace{-4pt}}

\newcommand {\Exp} {\mathit{t}}
\newcommand {\Var} {\mathit{x}}
\newcommand {\Const} {\mathit{c}}
\newcommand {\case} {\mathbf{case}\ }
\newcommand {\of} {\ \mathbf{of}\ }
\newcommand {\clet} {\ \mathbf{let}\ }
\newcommand {\cin} {\ \mathbf{in}\ }
\newcommand {\empt} {()}
\newcommand {\suc} {\mbox{\sf{succ}}}
\newcommand {\id} {\mbox{\sf{id}}}
\newcommand {\apply} {\mbox{\sf{apply}}}
\newcommand {\fix} {\mbox{\sf{fix}}}
\newcommand {\apple} {\mbox{\sf{apple}}}

\newcommand {\banana} {\mbox{\sf{banana}}}
\newcommand {\pred} {\mbox{\sf{pred}}}
\newcommand {\add} {\mbox{\sf{add}}}
\newcommand {\cons} {\mbox{\sf{cons}}}
\newcommand {\map} {\mbox{\sf{map}}}
\newcommand {\double} {\mbox{\sf{double}}}

\newcommand {\append} {\mbox{\sf{append}}}
\newcommand {\reverse} {\mbox{\sf{reverse}}}
\newcommand {\fourier} {\mbox{\sf{fourier}}}
\newcommand {\phases} {\mbox{\sf{phases}}}
\newcommand {\cphase} {\textit{cR}\,}
\newcommand {\alice} {\mbox{\sf{alice}}}
\newcommand {\bob} {\mbox{\sf{bob}}}
\newcommand {\cZ} {\textit{cZ}\,}
\newcommand {\cX} {\textit{cX}\,}
\newcommand {\epr} {\mbox{\sf{epr}}}
\newcommand {\deutsch} {\mbox{\sf{deutsch}}}
\newcommand {\teleport} {\mbox{\sf{teleport}}}
\newcommand {\Had} {H}
\newcommand {\X} {X}
\newcommand {\Y} {Y}
\newcommand {\Z} {Z}
\newcommand {\phase} {S}
\newcommand {\R} {R_3}
\newcommand {\cnot} {\textit{cnot}\,}
\newcommand {\His}{\mathcal{H}}

\newcommand {\un} {\underline}
\newcommand {\ov} {\overline}
\newcommand {\ph} {\underline{\phantom{0}}}

\newcommand {\val}[1]{\langle\!\langle #1 \rangle\!\rangle}

\newcommand {\ncase}[4] {
\case #1 \of \left\{
 \begin{aligned}
    &\un0 \to #2 \\
    &\suc\  #3 \to #4
      \end{aligned}
  \right.
  }

\newcommand {\lcase}[4] {\!\begin{aligned}[t]
                         &\case #1 \of
                         \left\{
                          \begin{aligned}
                           &\empt \to #2   \\
                           &#3  \to #4  \\
                          \end{aligned}
                          \right.
                         \end{aligned}}

\begin{document}

\title{
       A Lambda Calculus for Quantum Computation}
    \author{Andr\'e van Tonder
            \\ \\
            Department of Physics, Brown University \\
            Box 1843, Providence, RI 02906 \\
            andre@het.brown.edu
            }
    \date{July 15, 2003 \\
          Revised version: March 24, 2004}

    \maketitle

    \begin{abstract}
        \noindent
        The classical lambda calculus may be regarded both as a programming language
        and as a formal algebraic system for reasoning about computation.
        It provides a computational model equivalent to the Turing machine, and
        continues to be  of
        enormous benefit in the classical theory of computation.
        We propose that
        quantum computation,
        like its classical counterpart, may benefit
        from a version of the lambda calculus suitable for expressing and
        reasoning about quantum algorithms.
        In this paper we develop a quantum lambda
        calculus as an alternative model of quantum computation, which
        combines some of the benefits of both the
        quantum Turing machine and the quantum circuit models.
        The calculus
        turns out to be closely related to the linear lambda calculi
        used in the study of Linear Logic.  We set up a computational model
        and an equational proof
        system for this calculus, and
        we argue that it is equivalent to the quantum Turing machine.

        \noindent
        \rule{4.6in}{1pt}
        \newline
        {Brown preprint:} BROWN-HET-1366
    \end{abstract}

\section{Introduction}

Currently there exist two main approaches to the theory of quantum
computation:  The
quantum Turing machine,  introduced  by Benioff  and
Deutsch \cite{qtm1, qtm2}, and the  quantum circuit model,
introduced by Deutsch  \cite{circuits}.  These two approaches
were shown to be essentially equivalent by Yao \cite{yao}.

The quantum Turing machine provides a fundamental model of
quantum computation that may be  regarded as a baseline for defining universality.
However, reasoning
about Turing machines can be a cumbersome process, requiring
word-at-a-time thinking while keeping track of complicated machine
and tape states.  Turing machine programs do not satisfy a simple
algebra.

For this reason, the quantum circuit model is
more popular in the practical investigation of quantum algorithms.
Quantum circuits are visual, compositional, and may be manipulated
algebraically.  However, no single finite quantum circuit is universal.
Indeed, Yao's proof of Turing equivalence relies on the concept of
uniform circuit families generated by classical computation \cite{yao, nielsen}.
To define what we mean by such a circuit family, we need
to rely on a separate model of classical computation not
described by any finite quantum circuit.

In classical computation, the {lambda calculus}
provides an alternative computational
model, equivalent to the Turing machine, which
continues to be  of
enormous utility in the theory of computation, in mathematical logic,
and in the study of computer languages
and their semantics \cite{barendregt, mitchell, pierce, davis, gunter}.
Due to its simplicity and expressive power, the lambda calculus has
been used as the basis
of several powerful computer languages, including
Lisp, Scheme, ML and Haskell \cite{lisp, epl, ml, haskell}.

In this article, we propose that
quantum computing,
like its classical counterpart, may benefit
from an alternative computational model based on a
version of the lambda calculus suitable for expressing and
reasoning about quantum algorithms.
We develop such a calculus, which
turns out to be closely related to the linear lambda calculi
used in the study of Linear Logic.  We set up its computational model
and equational proof
system, and
argue that the computational model is equivalent to the quantum Turing machine.

The quantum lambda calculus combines
some of the benefits of both quantum circuits and the quantum
Turing machine.
The quantum lambda calculus describes functions that may be composed
and manipulated algebraically,
like quantum circuits.  Programs can be algebraically transformed into equivalent programs, and one
can solve equations
whose unknowns are programs, in much the same way as one solves equations
in high school algebra \cite{backus}. Unlike quantum circuits,
the quantum lambda calculus provides a
unified framework that is
universal for quantum computation without the need
to rely on a separate model of classical computation.

In a practical vein, we show how various known quantum algorithms may be expressed
as simple programs in the lambda calculus.
Indeed, the calculi described in this paper may be
used as a programming language for prototyping quantum algorithms.
In fact,
the algorithms exhibited in this article were transcribed into Scheme
for testing.  The simulator, which was also written in Scheme,
is available upon request from the author.

Since the first version of this paper was written, some progress has
been made by the author in devising a typed version, with accompanying
denotational semantics, of a fragment of the
quantum calculus described here \cite{myself}.

\section{The classical lambda calculus}
\label{classical}

We begin by providing a reasonably self-contained introduction to concepts
and constructions in the classical lambda calculus
that will be used in the rest of the paper.  The intended audience for
this section includes physicists and general computer scientists.  The expert
may skip this section and refer back as needed.

The classical lambda calculus may be regarded both as a programming language
and as a formal algebraic system for reasoning about computation.
It was originally introduced by Church in the
study of the foundations of mathematics \cite{church1, church2}.
Church postulated that it provides a universal model of
computation, which was later shown by Turing to be equivalent to the Turing
machine \cite{turing}.

As a formal system, the lambda calculus has axioms and rules of
inference, and lends itself to analysis using the language and
tools of mathematical logic.
Computation may be regarded as guided deduction in this formal
system.  This provides a directed form of equational reasoning that
corresponds to symbolic evaluation of programs via a sequence
of algebraic simplifications called reductions
\cite{barendregt, mitchell, pierce, davis, gunter}.

The syntax of the classical untyped lambda calculus $\lambda$ is as follows:
Expressions (also called \textit{terms})
are constructed recursively
from variables $x$, $y$, $z$, \dots, parentheses, spaces, the period, and the symbol $\lambda$,
according to the grammar of figure \ref{lambdasyntax}.
\rulefig{
\begin{align*}
  \Exp ::= &{}  &&\textit{terms:} \\
      &\Var\  &&\textit{variable} \\
      &(\lambda \Var.\,\Exp)\   &&\textit{abstraction}   \\
      &(\Exp\ \Exp)  &&\textit{application}
\end{align*}
}{Syntax of the lambda calculus $\lambda$ \label{lambdasyntax}}

A term of the form $(\lambda x.\,\Exp)$ is called
a \textit{functional abstraction}.  It represents the function
$x \mapsto \Exp$.  For example, the identity function $x\mapsto x$ is written as
\bcode
(\lambda x.\,x)
\ecode
The dummy
variable $x$ here is called a bound variable, and conforms to the
usual rules governing bound variables in mathematical formulae.
For example, we identify expressions that differ only in the
renaming of bound variables.

A term of the form $(\Exp\ \Exp)$ represents
a \textit{function application}.
The sole means of computation in the lambda calculus
is the operation of applying a function to
its argument consistent with the following axiom:
\begin{align*}
&((\lambda x.\,\Exp)\ v) = \Exp\,[v/x] &&(\beta)
\end{align*}
Here $v$ denotes a \textit{value}, to be defined shortly.
Reading this axiom from left to right defines
an algebraic rewrite rule
for transforming terms,
substituting the argument $v$ in place of
the variable $x$ into the function body.  This transformation is called
\textit{beta reduction}.  We will use the arrow $\longto$ to indicate
one (and sometimes more than one) beta reduction step.
A reducible term is called a
\textit{redex}.

Unabridged lambda terms can be painful to read.  For this reason, we
will often introduce abbreviations using the symbol $\equiv$.
In addition, we will often omit parentheses according to the convention that nested lambda
abstractions associate to the right and applications associate to the left.

Consider the simple program $((\lambda x.\,x)\ \apple)$ where $\apple$
stands for some term in our language.
With the abbreviation $\id \equiv (\lambda x.\,x)$, this may be
written more legibly as $(\id\ \apple)$, which should evaluate to $\apple$.
Indeed, beta reduction gives
\bcode
((\lambda x.\,x)\ \apple) \longto \apple
\ecode
in a single step.

In general, a computation consists of a sequence of beta reductions
executed according to some deterministic
strategy until the resulting term cannot be reduced any
further, at which point the computation terminates.

A slightly more complicated example, which serves to show how
multiple-argument functions can represented in terms of nested
single-argument functions (a technique known as
\textit{currying}), is given by
\begin{align*}
\apply &\equiv \lambda f.\,\lambda x.\,(f\ x) \\
       &\equiv \lambda f.\,(\lambda x.\,(f\ x))
\end{align*}
which represents a function that applies its first argument
$f$, which should be a function, to its second argument
$x$.  Applying the identity function to $\banana$ should
give $\banana$.  To see this, the program
$(\apply\ \id\ \banana)$, which is shorthand for $((\apply\ \id)\ \banana)$, is now
executed by the following sequence of beta reductions (underlining redexes)
\begin{align*}
((\apply\ \id)\ \banana) &\equiv (\un{((\lambda f.\,(\lambda x.\,(f\ x)))\ \id)}\ \banana) \\
                    &\longto (\un{(\lambda x.\,(\id\ x))}\ \banana)  \\
                    &\longto \un{(\id\ \banana)}  \\
                    &\longto \banana
\end{align*}
Often there is more than one reducible subterm
at any given step and a strategy is required to
make the process unambiguous. For definiteness, we will use a
\textit{call by value} strategy.
This works as follows:
Abstractions (terms of the form $(\lambda x.\,\Exp)$) are
considered \textit{values} and may not be reduced any further.  A
function application $(\Exp\ \Exp)$ may only be reduced if both
the operator and the operand are values.  Otherwise the operator
and operand must be reduced first.  We will call the
resulting calculus the call-by-value lambda calculus $\lambda_v$.
Formally, we state the syntax for values \cite{callbyvalue0, callbyvalue1}
in figure \ref{valuecbv}.
\rulefig{
\begin{align*}
v ::= &{} &&\textit{values:} \\
      &x &&\textit{variable} \\
      &(\lambda x.\,\Exp) &&\textit{abstraction value}
\end{align*}
}{Values in the call-by-value calculus $\lambda_v$ \label{valuecbv},
}

\noindent
The reduction rules are listed in figure \ref{reducecbv}.
\rulefig{
\brules
\begin{align*}
{t_1 \longto t_1' \over
 (t_1\ t_2) \longto (t_1'\ t_2)} & &&(\textit{app}_1)\\
{t_2 \longto t_2' \over
 (v_1\ t_2) \longto (v_1\ t_2')} & &&(\textit{app}_2)\\
(\lambda x.\,t)\ v \longto t\,[v/x]
& &&(\beta)
\end{align*}
\erules
}
{Reduction rules for the call-by-value calculus $\lambda_v$ \label{reducecbv}
}

\newpage
\noindent
We will denote by $\val{\Exp}$ the term, when it exists,
obtained by fully reducing $\Exp$ to a value.

We will often use a less cumbersome informal notation when defining functions.  For
example, the $\apply$ function above satisfies the following property
\bcode
\apply\ f\ x \longto (f\ x)
\ecode
under beta reduction.
Given this specification,
the translation into a lambda term is straightforward.

How do we represent data in the lambda calculus?  Since all we have at our
disposal are lambda terms, we need a way of encoding
data as lambda abstractions with specified properties.
There is a technique which can be used for any
kind of data structure, which we will illustrate with two examples:
Natural numbers and lists.

Let us first consider how the natural numbers may be represented.
As with any kind of data, we need a way to construct natural numbers
and a way to deconstruct them, extracting their constituents.

One possible encoding is as the
sequence
\bcode
\un0,\quad \un1 \equiv \val{\suc\  \un0},
\quad \un2 \equiv \val{\suc\  \un1},\quad \dots,
\ecode
where
\begin{align*}
\un0 &\equiv \lambda x.\,\lambda y.\,(x\ \id) \\
\suc &\equiv \lambda n.\,\lambda x.\,\lambda y.\,(y\ n)
\end{align*}
are the constructors.\footnote{Explicitly
\begin{align*}
0 &\equiv \lambda x.\,\lambda y.\,(x\ (\lambda w.\,w)) \\
1 &\equiv \lambda x.\,\lambda y.\,(y\ \lambda x.\,\lambda y.\,(x\ (\lambda w.\,w))) \\
2 &\equiv \lambda x.\,\lambda y.\,(y\ \lambda x.\,\lambda y.\,(y\  \lambda x.\,\lambda y.\,(x\ (\lambda w.\,w)))) \\
\vdots
\end{align*}
}
The above definitions were motivated by the need to be able
to define a $\case$ expression (deconstructor)
\bcode
\case \Exp_1 \of (\un0 \to \Exp_2,\ \suc\  m \to \Exp_3)
\ecode
which may now be taken as an abbreviation for
\bcode
\Exp_1\ (\lambda z.\,\Exp_2)\ (\lambda m.\,\Exp_3)
\ecode
Here $z$
denotes a variable that does not appear free in $\Exp_2$.  This
expression allows us to deconstruct a natural number, extracting the ingredients that went
into its construction, i.e., either $0$ or its predecessor $m$.
It is indeed not difficult to verify the following behavior under beta reduction:
\begin{align*}
\case \un0 \of (\un0 \to \Exp_2,\, \suc\  m \to \Exp_3)
    &\longto \Exp_2 \\
\case \val{\suc\  \Exp_0} \of (\un0 \to \Exp_2,\, \suc\  m \to \Exp_3)
    &\longto \Exp_3\,[\val{\Exp_0}/m]
\end{align*}
As an example, it is now trivial to define the predecessor function (with the
convention that $\pred\ \un0 = \un0$)
\bcode
\pred \equiv \lambda n.\,\ncase {n} {\un0}
                               {m} {m}
\ecode
In order to program arbitrary computations, we need to verify that
the lambda calculus is sufficiently powerful to
represent \textit{recursive} functions.  Indeed, recursion can be used
to represent any kind of iterative or looping computation.

For
example, in order to define addition, we need an expression $\add$
which behaves as follows under beta reduction:
\bcode
\add\  m\ n \longto \ncase {m}
                        {n}
                        {k} {\add\  k\ (\suc\  n)}
\ecode
where the subterm denoted by $\add$ on the left has copied itself into the
body of the term on the right hand side.  One of the simplest ways of achieving this
is to define \cite{oleg}
\begin{equation}
  \add \equiv (t\ t)
  \label{recurs1}
\end{equation}
where
\begin{equation}
  t \equiv
     \lambda f.\left(\lambda m.\,\lambda n.\,
                           \ncase {m}
                                  {n}
                                  {k} {(f\ f)\ k\ (\suc\  n)}\right)
  \label{recurs2}
\end{equation}
In other words, $t$ is an abstraction consisting of the body of the addition
function with the combination $(f\ f)$ in the position where $\add$ should
insert itself after reduction.
It is a simple exercise to show that $\add$ indeed has the specified behavior under
beta reduction.  This method can be applied to any recursive function.

The computation of the program $(\add\  2\ 2)$ then proceeds via the following sequence
of beta reductions:
\begin{align*}
\add\  \un2\ \un2 &\equiv \add\  \val{\suc\  \un1}\ \un2  \\
          &\longto \case \val{\suc\  \un1} \of (\un0 \to \un2,\, \suc\  k \to \add\  k\ (\suc\  \un2)) \\
          &\longto \add\  \un1\ \val{\suc\  \un2} \equiv \add\  \val{\suc\  \un0}\ \un3 \\
          &\longto \case \val{\suc\  \un0} \of (\un0 \to \un3,\, \suc\  k \to \add\  k\ (\suc\  \un3)) \\
          &\longto \add\  \un0\ \val{\suc\  \un3} \equiv \add\  \un0\ \un4 \\
          &\longto \case \un0 \of (\un0 \to \un4, \suc\  k \to \add\  k\ (\suc\  \un4)) \\
          &\longto \un4
\end{align*}
The above technique can be generalized to arbitrary data structures.  For example,
lists can be represented by the following constructors, which are entirely
analogous to those of the natural numbers
\begin{align*}
\empt &\equiv \lambda x.\,\lambda y.\,(x\ \id) \\
\cons &\equiv \lambda h.\,\lambda t.\,\lambda x.\,\lambda y.\,((y\ h)\ t)
\end{align*}
where $\empt$ denotes the empty list
and $\cons$ constructs a list consisting of a first (head) element $h$
followed by a list $t$ (the tail) containing the rest of the elements.
Again, the above definitions were motivated by the need
to be able to define a deconstructor
\bcode
\case \Exp_1 \of (\empt \to \Exp_2,\, \cons\ h\ t \to \Exp_3)
\ecode
which may now be taken as an abbreviation for
\bcode
\Exp_1\ (\lambda z.\,\Exp_2)\ (\lambda h.\,\lambda t.\,\Exp_3)
\ecode
We will often abbreviate $h:t \equiv (\cons\ h\ t)$.  We can define
tuples in terms of lists as
$(x_1, \ldots, x_n) \equiv {x_1:x_2:\cdots :x_n:\empt}$.
Under beta reduction, we have the behavior
\begin{align*}
\case \empt \of (\empt \to \Exp_2,\, h:t \to \Exp_3)
      &\longto \Exp_2 \\
\case \val{\Exp_0:\Exp_1} \of (\empt \to \Exp_2,\, h:t \to \Exp_3)
      &\longto \Exp_3\,[\val{\Exp_0}/h,\, \val{\Exp_1}/t]
\end{align*}
showing how the case expression may be used to deconstruct the list, extracting its
head and tail.
To see how these abstractions are used, consider the following
recursive function
\bcode
\map\ f\ \mathit{list} \longto
          \lcase {\mathit{list}}
                 {\empt}
                 {h:t} {(f\ h):(\map\ f\ t)}
\ecode
which takes as input a function and a list and applies the function
to each element of the list.  The reader may verify that, for example,
\bcode
\map\ \double\ (\un4,\, \un7,\, \un2) \longto (\un8,\, \un{14},\, \un4), \qquad \double \equiv
\lambda x.\, (\add\  x\ x)
\ecode
Finally, we introduce some convenient notation.  Since we can represent tuples as lists,
we can define functions on tuples using notation such as
\bcode
\lambda(x, y).\,\Exp \equiv \lambda u.\,
                   \lcase {u}   {\empt}
                           {x:t'} {\lcase {t'} {\empt}
                                         {y:t''} {\Exp}}
\ecode
For example, $(\lambda(x, y).\,(\add\  x\ y))\ (\un7,\, \un7)$ evaluates to $\un{14}$.
It is also useful to have a notation for representing intermediate results.
The $\clet$ notation
\begin{align*}
&\clet (\Var_1,\ldots, \Var_n) = (\Exp_1, \ldots, \Exp_n) \cin \Exp  \\
    &\qquad\quad\equiv (\lambda(\Var_1, \ldots, \Var_n).\,\Exp)\ (\Exp_1, \ldots, \Exp_n)
\end{align*}
allows us to write terms such as
\bcode
\begin{aligned}
&\clet x = \un1 \cin  \\
&\clet (y, z) = (\un2, \un3) \cin \\
&\ \ \add\  x\ (\add\  y\ z)
\end{aligned}
\ecode
which evaluates to $\un6$.

\section{A quantum computational model}

In this section we will construct a computational model, based on
the lambda calculus, suitable for describing quantum computations.
The language used will be an adaptation of the classical lambda calculus,
extended with a set of quantum primitives.  We will denote it by
$\lambda_i$ where the subscript stands for \textit{intermediate}.
For reasons to be discussed in the next section, this language
is not suitable as a formal system.  In particular, reduction in $\lambda_i$
does not correspond to a simple system for equational
reasoning.  In section \ref{sqlambda} we will correct these deficiencies to obtain the
full quantum lambda calculus $\lambda_q$.

In the classical lambda calculus, beta reduction consumes the
program to give the result.  At each step, information is discarded, which
makes the process irreversible.
For quantum computing, we need reduction rules that
take computational states to superpositions of states in a way
that is unitary and reversible.

Bennett \cite{bennettrev} showed that any
classical computation can be transformed into a reversible computation.
The construction, adapted to our situation,
is as follows:  Let $x$ denote the term being computed,
and let $\beta:x \mapsto \beta(x)$ denote a single beta reduction step.
Instead of the non-invertible function $\beta$, one considers the
function $x \mapsto (x, \beta(x))$, which is invertible on its
range.  In its
simplest version, the computation proceeds as
$$
  x \mapsto (x, \beta(x)) \mapsto (x, \beta(x), \beta^2(x))
    \mapsto (x, \beta(x), \beta^2(x), \beta^3(x)) \mapsto \cdots
$$
More complicated schemes exist that reversibly erase the intermediate steps,
saving space at the expense of running time.
Although this process does not end by itself, we may observe it and
regard the computation as having terminated when
$\beta^{n+1}(x) = \beta^n(x)$, at which time we may stop the machine by
external intervention.

Although this scheme can be used to reversibly implement computations
in the classical lambda calculus, we will soon see that it does not work
unmodified in the quantum case.

In order to represent computations involving qubits, we will add a few constant
symbols as additional
primitives to our language as in figure \ref{syntaxi}.
\rulefig{
\begin{align*}
  \Exp ::= &{}  &&\textit{terms:} \\
      &\Var\  &&\textit{variable} \\
      &(\lambda \Var.\,\Exp)\   &&\textit{abstraction}   \\
      &(\Exp\ \Exp)  &&\textit{application} \\
      &\Const  &&\textit{constant} \\
  \Const ::= &{} &&\textit{constants:} \\
    &0\ |\ 1\ |\ \Had\ |\ \phase\ |\ \R\ |\ \cnot\ |\ \X\ |\ \Y\ |\ \Z\ |\ \dots
\end{align*}
}{
Syntax of the intermediate language $\lambda_i$  \label{syntaxi}
}

The symbols $0$ and $1$ here are \textit{primitives}
and should not be confused with the abbreviations
$\un0$ and $\un1$ of the previous
section.  Additional constants $\Had$, $\phase$, \dots, will denote elementary
gate operations on qubits.  These should include symbols for a universal set of elementary
quantum gates \cite{circuits, gates1, gates2}.  For example,
the set consisting of the Hadamard gate $\Had$, the phase gate $\phase$, the
$\pi/8$ gate $\R$, and the controlled-not gate $\cnot$ is universal
\cite{gates2, nielsen}.
Additional primitives, such as the the Pauli gates $\X$, $\Y$ and $Z$,
may be added for convenience.

We now allow the state of a computation to be a quantum superposition
of terms in this language.  As a model, one
 may imagine lambda terms encoded as strings of
symbols on the tape of a quantum Turing machine.

As a first example, consider an initial state written in ket notation as
$$
  \ket {(\Had\ 0)}.
$$
We would like to choose the transition
rules of the quantum computer in such a way that this string will
evaluate to the Hadamard operator applied to
$\ket{0}$, which should give the superposition $\rthalf \left(\ket{0} +
\ket{1}\right)$ of the states $\ket{0}$ and $\ket{1}$ containing
unit-length strings.  The candidate reduction rule
\begin{align*}
 \ket {(\Had\ 0)} &\longrightarrow
  \rthalf \left(\ket{0} + \ket{1}\right) \\
  \ket {(\Had\ 1)} &\longrightarrow
  \rthalf \left(\ket{0} - \ket{1}\right)
\end{align*}
is not reversible.  To make it reversible, we first try the same trick
as in the classical case
\begin{align*}
 \ket {(\Had\ 0)} \longrightarrow
  &\rthalf \left(\ket{(\Had\ 0) ;\, 0} + \ket{(\Had\ 0) ;\, 1}\right) \\
  =
  &\ket{(\Had\ 0)}\otimes\rthalf \left(\ket{0}
  + \ket{1}\right)
\end{align*}
where we have factored out the common substring.  The semicolon denotes
string concatenation.  In this simple example, the
answer indeed factors out on the right.  However, notice what happens if
we apply this method to the term
\begin{align*}
 \ket {(\Had\ {(\Had\ 0)})}
 &\longto \rthalf\left(
   \ket{(\Had\ (\Had\ 0)) ;\, {(\Had\ 0)}}
      + \ket{(\Had\ (\Had\ 0)) ;\, {(\Had\ 1)}}\right) \\
   &\longto \half
  \ket{(\Had\ (\Had\ 0))}\otimes \\
  &\qquad\quad\otimes\biggl(\ket{(\Had\ 0) ;\, 0}
      + \ket{(\Had\ 0) ;\, 1}
      + \ket{(\Had\ 1) ;\, 0}
      - \ket{(\Had\ 1) ;\, 1}\biggr)
\end{align*}
Here the answer does not factor out.  The fully reduced
rightmost term is entangled with the intermediate
term in the history.

Note, however, that this scheme keeps more information than necessary.
For reversibility, it is sufficient to record at each step only
which subterm has been reduced, and the operation that has been
applied to it.  We may encode this in our example as follows (to be formalized below):
\begin{align*}
 \ket {(\Had\ {(\Had\ 0)})}
 \longto &\rthalf\left(
   \ket{(\ph\ (\Had\ \ph)) ;\, {(\Had\ 0)}}
      + \ket{(\ph\ (\Had\ \ph)) ;\, {(\Had\ 1)}} \right) \\
   \longto &\half
    \ket{(\ph\ (\Had\ \ph))}\otimes \\
  &\quad\otimes\biggl(\ket{(\Had\ \ph) ;\, 0}
      + \ket{(\Had\ \ph) ;\, 1}
      + \ket{(\Had\ \ph) ;\, 0}
      - \ket{(\Had\ \ph) ;\, 1}\biggr) \\
  =
  &\ket{(\ph\ (\Had\ \ph)) ;\, (\Had\ \ph)}\otimes
   \ket{0}
\end{align*}
Here we have at each step replaced subterms that do not need to be
recorded
 by the constant placeholder symbol $\ph$.  Now the answer does indeed factor
 out on the right as required, consistent with $H^2 \ket 0 = \ket 0$.
It is also clear that at each step we have kept enough information to
reconstruct the previous step, thus ensuring reversibility.

\rulefig{
\begin{align*}
v ::= &{} &&\textit{values:} \\
      &x &&\textit{variable}\\
      &\Const &&\textit{constant} \\
      &(\lambda x.\,\Exp) &&\textit{abstraction value}
\end{align*}
}{
Values in the intermediate language $\lambda_i$ \label{iconstants}
}

The computational model may now be formalized with the following rules:  First,
we extend our definition of values to include constants as in figure \ref{iconstants}.
The computational state is taken to be a quantum superposition of sequences of
the form
$$
  h_1;\,\ldots;\, h_n;\,t
$$
where $h_1;\,\ldots;\,h_n$ will be called the \textit{history track} and $t$
will be called
the \textit{computational register}.  The classical subset
of the transition rules is shown in figure \ref{ioperational}.
\rulefig{
\brules
\begin{align*}
&\!\!\!\!\!\!\!\!\!\!\!\!\!\!\!\!\!\!\!\!\!\!\!\!\!
  {\!\!\!\!\!\!\!\!\!t_1 \longto h_1;\,t_1' \over
 \His;\,(t_1\ t_2) \longto \His;\,(h_1\ \ph);\, (t_1'\ t_2)}  &&(\textit{app}_1)\\
&\!\!\!\!\!\!\!\!\!\!\!\!\!\!\!\!\!\!\!\!\!\!\!\!\!\!
{\!\!\!\!\!\!\!\!\!t_2 \longto h_2;\,t_2' \over
 \His;\,(v_1\ t_2) \longto \His;\,(\ph\ h_2);\,(v_1\ t_2')}  &&(\textit{app}_2) \\
\His;\,((\lambda x.\, \Exp)\ v) &\longto \His;\,((\lambda x.\, {\ov\Exp}_x)\ \ph);\,\Exp\,[v/x]
    &&\text{($\beta_1$) if $x$ appears free in $\Exp$ } \\
\His;\,((\lambda x.\, \Exp)\ v) &\longto \His;\,((\lambda x.\, \ph)\ v);\,\Exp
                        &&\text{($\beta_2$) if $x$ not free in $\Exp$ } \\
  \His;\, t &\longto \His;\,\ph;\,t   &&\text{(\textit{Id}) otherwise }
\end{align*}
\erules
}{
Operational model for the classical subset of $\lambda_i$ \label{ioperational}
}

In these rules $\His$ denotes the (possibly empty) history track, and
 ${\ov \Exp}_x$ is obtained from $\Exp$ by
recursively replacing all subterms that do not contain $x$ with
the placeholder symbol and keeping $x$.  More formally
\begin{equation}
\begin{aligned}
   \ov {t}_x &\equiv \ph &&\text{if $x$ not free in $t$} \\
   \ov {(\lambda y.\,\Exp)}_x &\equiv (\ph.\,\ov{\Exp}_x)
    \\
   \ov {(\Exp\ \Exp')}_x &\equiv (\ov{\Exp}_x\ \ov{\Exp'}_x) \\
   {\ov x}_x &\equiv x
\end{aligned} \label{overl}
\end{equation}
These rules are sufficient to make classical
computations reversible, provided that lambda terms
which differ only by renaming of bound
variables have been identified.  In this regard, we note here
that in a quantum Turing machine model, it is possible to represent terms
on the tape of the quantum Turing machine in an unambiguous way
(e.g., using De Bruijn indices instead of bound variables)
\cite{mitchell, pierce}.

\noindent
Here is an example computation:
\begin{align*}
&\ket{((\apply\ \id)\ \banana)} \equiv ({((\lambda f.\,(\lambda x.\,(f\ x)))\
     (\lambda z.\,z))}\ \banana) \\
     &\qquad
      \begin{aligned}[t]
                    &\longto \ket{(((\lambda f.\,(\ph.\,(f\ \ph)))\ \ph)\ \ph)
                    ;\,{((\lambda x.\,((\lambda z.\,z)\ x))\ \banana)}}  \\
                    &\longto \ket{(((\lambda f.\,(\ph.\,(f\ \ph)))\ \ph)\ \ph)
                    ;\,((\lambda x.\,(\ph\ x))\ \ph)
                    ;\,((\lambda z.\,z)\ \banana)}  \\
                    &\longto \ket{(((\lambda f.\,(\ph.\,(f\ \ph)))\ \ph)\ \ph)
                    ;\,((\lambda x.\,(\ph\ x))\ \ph)
                    ;\,((\lambda z.\,z)\ \ph)
                    ;\,\banana
                    }  \\
                    &\longto \ket{(((\lambda f.\,(\ph.\,(f\ \ph)))\ \ph)\ \ph)
                    ;\,((\lambda x.\,(\ph\ x))\ \ph)
                    ;\,((\lambda z.\,z)\ \ph)
                    ;\,\ph
                    ;\,\banana
                    }  \\
                    &\longto \ket{(((\lambda f.\,(\ph.\,(f\ \ph)))\ \ph)\ \ph)
                    ;\,((\lambda x.\,(\ph\ x))\ \ph)
                    ;\,((\lambda z.\,z)\ \ph)
                    ;\,\ph
                    ;\,\ph
                    ;\,\banana
                    }
\\
                    &\longto \cdots
      \end{aligned}
\end{align*}
At each step just enough information is kept to reconstruct the previous step.
Although in this particular example, termination can be tested by
observing and comparing the last expression in the history with $\ph$,
in general we do not have a
well-defined criterion for termination in the calculus $\lambda_i$, due
to the fact that
the state may involve a superposition of several computational
histories, some of which have terminated and others not. Thus, to
observe termination would potentially disturb the state. This
problem will be solved in the quantum calculus $\lambda_q$ of section
\ref{sqlambda}.

In addition, we have some extra reduction rules involving the quantum
gate symbols such as:
\rulefig{
\brules
\begin{align*}
  \ket{\His;\,(\Had\ 0)} &\longto \ket{\His;\,(\Had\ \ph)}\otimes\rthalf\left(\ket 0 + \ket 1\right) \\
  \ket{\His;\,(\Had\ 1)} &\longto \ket{\His;\,(\Had\ \ph)}\otimes\rthalf\left(\ket 0 - \ket 1\right)
\end{align*}
\erules
}{
Operational model for $\Had$
}

\noindent
The rules for quantum primitives are summarized in figure \ref{iquantum}.
\rulefig{
\brules
\begin{align*}
  \ket{\His;\,(c_U\ \phi)} \longto \ket{\His;\,(c_U\ \ph)}\otimes U\ket \phi
      &&(U)
\end{align*}
\erules
}{
Operational model for the quantum primitives of $\lambda_i$ \label{iquantum}
}

\noindent
Here $c_U$ denotes any one of the quantum primitive symbols and $U$ the
corresponding unitary transformation, while $\phi$ stands for $0$ or $1$
in the case of single-bit operators, or one of $(0, 0)$, $(0, 1)$, $(1, 0)$ or
$(1, 1)$ in the case of two-bit operators.  For example
$$
  \ket{(\cnot\ (1, 0))} \longto \ket{(\cnot\ \ph);\, (1, 1)}.
$$

\section{Towards an equational theory}

While the language $\lambda_i$ constructed
in the previous section can be used to describe quantum
computations,
reduction in $\lambda_i$ does not correspond to a simple system
for equational reasoning.  This makes $\lambda_i$ unsuitable
as a formal proof system for quantum computation.
We will discuss the problem
in this section and resolve it in the next with the introduction of
 the quantum lambda calculus $\lambda_q$.

In the classical lambda calculus, program evaluation through
beta reduction can be regarded as
a directed form of equational reasoning consistent with the axiom
\begin{align*}
(\lambda x.\, t)\ v = t\,[v/x] &&(\beta)
\end{align*}
Indeed, the classical lambda calculus provides
both a model of computation and a formal system for
reasoning about functions, a property we would like to keep
in the quantum case.

To understand the difficulty,
notice what happens when a function application discards
its argument (in other words, the argument does not appear in the
function body).  For example
\begin{align*}
  \ket{((\lambda x.\,\apple)\ \banana)}
  &\longto \ket{((\lambda x.\,\ph)\ \banana);\, \apple}
\end{align*}
We see that in order to maintain reversibility, a record of
the argument $\banana$ is kept in the history.  Restricting our attention to the computational
register, we see that its evolution is consistent with replacing
the original expression with an equal expression according to the axiom ($\beta$).
In other words, in this example reduction is consistent with
equational reasoning.

However,
we run into problems when the discarded subterm is in a quantum superposision
with respect to the computational basis.  For example, consider the
reduction of
\begin{align*}
\ket{(\lambda x.\, 0)\ (\Had\ 0)}
&\longto
\ket{(\ph\ (\Had\ \ph)}\otimes \rthalf
\biggl(\ket {(\lambda x.\, 0)\ 0} + \ket {(\lambda x.\, 0)\ 1}\biggr)  \\
&\longto
\ket{(\ph\ (\Had\ \ph)}\otimes \rthalf
\biggl(\ket {(\lambda x.\,\ph)\ 0} + \ket {(\lambda x.\, \ph)\ 1}\biggr)
\otimes \ket 0
\end{align*}
In the second step, a discarded subterm in a superposition
is saved in the history and the computational register becomes
$\ket0$.  However, if we were to apply the axiom ($\beta$) to the contents
of the computational register, we would get the equation
$$
  \rthalf
\biggl(\ket {(\lambda x.\, 0)\ 0} + \ket {(\lambda x.\, 0)\ 1}\biggr)
  = \sqrt 2\,\ket 0
$$
which is invalid since the right hand side is not a legal normalized state.

As a second example, consider the following computation, where the inner function
discards its argument $x$:
\begin{align*}
&\ket{((\lambda y.\,((\lambda x.\,y)\ y))\ (\Had\ 0))} \\
 &\qquad\begin{aligned}[t]
   &\longto \rthalf
     \ket{(\ph\ (\Had\ \ph))}\otimes\biggl(\ket{((\lambda y.\,((\lambda x.\,y)\ y))\ 0)}
     + \ket{((\lambda y.\,((\lambda x.\,y)\ y))\ 1)}
     \biggr) \\
   &\longto \rthalf
     \ket{(\ph\ (\Had\ \ph));\,((\lambda y.\,((\ph.\,y)\ y))\ \ph)}
       \otimes \\
       &\phantom{\longto}\qquad \otimes
         \biggl(
          \ket{((\lambda x.\,0)\ 0)}
          +\ket{
          ((\lambda x.\,1)\ 1)} \biggr)\\
   &\longto \rthalf
     \ket{(\ph\ (\Had\ \ph));\,((\lambda y.\,((\ph.\,y)\ y))\ \ph)}
       \otimes \\
       &\phantom{\longto}\qquad \otimes
         \biggl(
          \ket{((\lambda x.\,\ph)\ 0);\, 0}
          +\ket{
          ((\lambda x.\,\ph)\ 1);\, 1} \biggr)
 \end{aligned}
\end{align*}
Now the computational register is entangled with the last expression in the
history.  Ignoring the history, the computational register
would be in a mixed state with density matrix
$$
\left(\begin{array}{cc}
\hhalf & 0 \\
0 & \hhalf
\end{array}\right)
$$
However, an attempt to apply the equational axiom ($\beta$)
to the contents of the computational
register would give
$$
\rthalf \biggl(\ket{((\lambda x.\,0)\ 0)}
          +\ket{
          ((\lambda x.\,1)\ 1)}\biggr)
 = \rthalf \left(\ket0 + \ket1\right)
$$
which is clearly inconsistent.

\section{A quantum lambda calculus}
\label{sqlambda}

We will resolve the shortcomings of the language $\lambda_i$ by
developing a quantum lambda calculus $\lambda_q$ which has a consistent
equational theory.  This section will be somewhat heavier on
the formalities, and the reader who wishes to see some
concrete examples may wish to skip ahead to section \ref{sexamples}
after reading the introductory paragraphs.

The previous discussion suggests that the problems with equational
reasoning in the presence of quantum operations can be avoided
by preventing functions from discarding arguments which
may be in a superposition with respect to the computational basis.

Let us call a subexpression \textit{definite} with
respect to the computational basis if it is textually the same in all branches of
the superposition. For example, in the state
$$
  \rthalf
\biggl(\ket {(\lambda x.\, 0)\ 0} + \ket {(\lambda x.\, 0)\ 1}\biggr)
$$
the subexpression $(\lambda x.\, 0)$ is definite, whereas the argument
$\half\left(\ket{0} + \ket{1}\right)$ is non-definite.
Definite subexpressions may be thought of as a classical resource.
They can be observed without affecting the state of the computation.
On the other hand, non-definite subexpressions represent
purely quantum resources.

To avoid the problems pointed out in the previous section, we
seek a calculus that will keep track of whether an
argument is definite or non-definite, and which will make it impossible to
write a function that discards a non-definite resource.
Calculi that are resource sensitive,
known as \textit{linear} lambda calculi, have
been studied intensively in recent years \cite{abramsky, linearsyn, linearcall, seely}.
So-called \textit{typed}
linear lambda calculi are very closely related to the field of
{linear logic} \cite{girard, lineartaste}.
Linear logic is a resource sensitive logic where, for example,
certain assumptions may only be used once in the course of a derivation.

For our purposes it will be sufficient to study a simple \textit{untyped}
linear calculus.
The syntax is a fragment of the one introduced in
\cite{linearsyn}, extended with quantum operations as in figure \ref{qsyntax}.
\rulefig{
\begin{align*}
  \Exp ::= &{}  &&\textit{terms:} \\
      &\Var\  &&\textit{variable} \\
      &(\lambda \Var.\,\Exp)\   &&\textit{abstraction}   \\
      &(\Exp\ \Exp)  &&\textit{application} \\
      &\Const  &&\textit{constant} \\
      &!\Exp &&\textit{nonlinear term}\\
      &(\lambda !\Var.\,\Exp)\   &&\textit{nonlinear abstraction}   \\
  \Const ::= &{} &&\textit{constants:} \\
    &0\ |\ 1\ |\ \Had\ |\ \phase\ |\ \R\ |\ \cnot\ |\ \X\ |\ \Y\ |\ \Z\ |\ \dots
\end{align*}
}{
Syntax of the quantum calculus $\lambda_q$ \label{qsyntax}
}

Here terms of the form $!t$ are called \textit{nonlinear}.  Nonlinear
terms will be guaranteed to be definite with respect to the computational
basis, and may be thought of physically as classical strings of symbols
that may be discarded and duplicated at will.  On the other hand,
linear terms may be non-definite, possibly containing embedded qubits
in superpositions with respect to the computational basis.
Abstractions of the form $(\lambda !\Var.\,\Exp)$ denote functions of
nonlinear arguments.  In an abstraction of the form $(\lambda \Var.\,\Exp)$,
the argument is called \textit{linear}.

A functional abstraction may use a nonlinear argument any number of times in its body,
or not at all.  On the other
hand, a linear argument must appear exactly once in the function body
(hence the name \textit{linear}).

To enforce these rules, we require terms to be \textit{well-formed}.
This corresponds to the constraint that linear arguments appear
linearly in a function body, and that all free variables appearing
in a term $!t$ refer to nonlinear variables \cite{linearcall}.  In the following
examples, the terms in the left column are well-formed, while those in the
right column are ill-formed
\footnote{
Notice that while well-formedness guarantees that linear resources will
be used appropriately, it does not guarantee that terms are
meaningful.  For example the term $(\lambda y.\, (\lambda\, !z.0)\  y)$
is well-formed, but may or may not get stuck at run-time, according to the
operational model of figure \ref{qoperational},
when applied to a linear or nonlinear argument respectively.  A typed
calculus would be needed to specify which terms can be legally
substituted for $y$.  For recent progress in this direction, see \cite{myself}.
}
\bcode
\begin{aligned}[l,l]
  &(\lambda !x.\,0)    &     &(\lambda x.\,0)  \\
  &(\lambda x.\,x)      &    &(\lambda x.\,!x)  \\
  &(\lambda !x.\,(x\ x))  &   &(\lambda x.\,(x\ x)) \\
  &(\lambda y.\,(\lambda !x.\,y))& &(\lambda y.\,(\lambda x.\,y)) \\
  &(\lambda !y.\,!(\lambda !x.\,y))&  &(\lambda y.\,!(\lambda !x.\,y)) \\
\end{aligned}
\ecode
Well-formedness
is a property that can be checked syntactically.
For completeness, we formally state the rules for well-formedness \cite{linearcall}, which
the reader satisfied with the above informal characterization may skip, in figure
\ref{well} .
\rulefig{
\brules
\begin{align*}
{{}\over{\vdash c}} & &&\textit{Const} \\
{{}\over{x \vdash x}} & &&\textit{Id} \\
{!x_1, \ldots, !x_n \vdash t \over !x_1, \ldots, !x_n \vdash !t} & &&\textit{Pomotion}\\
{\Gamma, x \vdash t \over \Gamma, !x \vdash t} & &&\textit{Dereliction} \\
{\Gamma, !x, !y \vdash t \over \Gamma, !z \vdash t\,[z/x, z/y]}
    & &&\textit{Contraction}  \\
{\Gamma \vdash t \over \Gamma, !x \vdash t} & &&\textit{Weakening} \\
{\Gamma, x \vdash t \over \Gamma \vdash (\lambda x.\,t)} &  &&\textit{$\multimap$-I}  \\
{\Gamma, !x \vdash t \over \Gamma \vdash (\lambda !x.\,t)}&  &&\textit{$\rightarrow$-I}  \\
{\Gamma \vdash t_1 \quad \Delta \vdash t_2\over
   \Gamma, \Delta \vdash (t_1\ t_2)}
    & &&\textit{$\multimap$-E}
\end{align*}
\erules
}{
Rules for well-formed terms in the quantum calculus $\lambda_q$ \label{well}
}

These rules may be related to the typed linear calculi described
in \cite{linearcall, linearsyn} by erasing the type annotations from the typing rules
of the latter.  Here $\Gamma$ and $\Delta$ denote \textit{contexts}, which are
sets containing linearity assumptions of the form $x$ and $!x$, where each variable $x$
is distinct.  If $\Gamma$ and $\Delta$ are contexts with no
variables in common, then $\Gamma, \Delta$
denotes their union.  For example, the rule
\textit{$\multimap$-E} implicitly assumes that $\Gamma\cap\Delta = \emptyset$.
Rules may be read as follows:  For example, the promotion
rule says that if $t$ is a well-formed term under the assumption that
$x_1$ to $x_n$ are nonlinear, then $!t$ is a well-formed term under the same assumption.
The condition $\Gamma \cap\Delta = \emptyset$ in ($\multimap$-E) ensures that a linear
variable can only appear once in the body of a formula.  The weakening and
($\rightarrow$-I) rules allow a function to discard a nonlinear argument, whereas
the contraction and ($\multimap$-E) rules allow us to duplicate a nonlinear argument
any number of times in the body of a function.

The well-formedness constraint prevents us from writing a function
which discards a linear argument.  However,
this is not sufficient to prevent unsafe computations without further
specification of the substitution order.  To see this,
consider the expression $((\lambda !x.\,0)\ !(\Had\ 0))$,
which is well-formed.  The problem is that we are allowed to use
$!$ to promote the expression $(\Had\ 0)$ to a nonlinear value, which
can then be discarded.
If we were allowed to reduce the subterm $(\Had\ 0)$ first,
equational reasoning would give
\begin{align*}
  \ket{((\lambda !x.\,0)\ !(\Had\ 0))}
     &= \rthalf\biggl( \ket{((\lambda !x.\,0)\ !0)} +
                             \ket{((\lambda !x.\,0)\ !1)} \biggr) \\
     &= \sqrt2 \ket0
\end{align*}
which is an invalid equation since
the last line is not a valid normalized state.  On the
other hand, if we consider $!(\Had\ 0)$ as an irreducible
value, we may use beta reduction
immediately to obtain
\begin{align*}
  \ket{((\lambda !x.\,0)\ !(\Had\ 0))}
     &= \ket0
\end{align*}
which is a valid result, since we are discarding the unevaluated expression
$!(\Had\ 0)$, which is definite.

To prevent terms of the form $!\Exp$ from being evaluated, we follow
Abramsky \cite{abramsky} and
extend our definition of values as in figure \ref{qvalues}.
\rulefig{
\begin{align*}
v ::= &{} &&\textit{values:} \\
      &x &&\textit{variable} \\
      &\Const &&\textit{constant} \\
      &(\lambda x.\,\Exp) &&\textit{linear abstraction} \\
      &(\lambda !x.\,\Exp) &&\textit{nonlinear abstraction} \\
      &!\Exp &&\textit{$!$-suspension}
\end{align*}
}{
Values in the quantum calculus $\lambda_q$ \label{qvalues}
}

\noindent
The computational model is described in figure \ref{qoperational}
\footnote{See \cite{abramsky, TurnerWadlerOperational, chirimar} for related operational
interpretations of linear lambda calculi.
Our evaluation model recomputes $!$-closures (see \cite{TurnerWadlerOperational}).},
where $\ov t$ is defined as in (\ref{overl}).
\rulefig{
\brules
\begin{align*}
&\!\!\!\!\!\!\!\!\!\!\!\!\!\!\!\!\!\!\!\!\!\!\!\!\!
  {\!\!\!\!\!\!\!\!\!t_1 \longto h_1;\,t_1' \over
 \His;\,(t_1\ t_2) \longto \His;\,(h_1\ \ph);\, (t_1'\ t_2)} &&(\textit{app}_1)\\
&\!\!\!\!\!\!\!\!\!\!\!\!\!\!\!\!\!\!\!\!\!\!\!\!\!\!
 {\!\!\!\!\!\!\!\!\!t_2 \longto h_2;\,t_2' \over
 \His;\,(v_1\ t_2) \longto \His;\,(\ph\ h_2);\,(v_1\ t_2')} &&(\textit{app}_2)\\
\His;\,((\lambda x.\, \Exp)\ v) &\longto \His;\,((\lambda x.\, {\ov\Exp}_x)\ \ph);\,\Exp\,[v/x]
   &&(\beta) \\
\His;\,((\lambda !x.\, \Exp)\ !t') &\longto \His;\,((\lambda !x.\, {\ov\Exp}_x)\ \ph);\,\Exp\,[t'/x]
   &&\text{($!\beta_1$) if $x$ appears free in $\Exp$ } \\
\His;\,((\lambda !x.\, \Exp)\ !t') &\longto \His;\,((\lambda !x.\, \ph)\ !t');\,\Exp
                        &&\text{($!\beta_2$) if $x$ not free in $\Exp$ } \\
  \ket{\His;\,(c_U\ \phi)} &\longto \ket{\His;\,(c_U\ \ph)}\otimes U\ket \phi
      &&(U)\\
  \His;\, t &\longto \His;\,\ph;\,t &&\text{(\textit{Id}) otherwise }
\end{align*}
\erules
}{
Operational model for the quantum lambda calculus $\lambda_q$ \label{qoperational}
}

According to these rules, quantum superpositions can only be created by evaluating terms
containing quantum primitives.   The result of applying a quantum gate
is a linear value, not preceded by a $!$.  As we prove below,
there is no way of including such a linear value in a nonlinear subterm.
It follows that subterms that may be
quantum non-definite will never be discarded
since, by ($!\beta_1$) and ($!\beta_2$),
nonlinear functions can only be applied to nonlinear terms.

Note that when a nonlinear function encounters a linear argument,
it simply gets stuck.  More precisely, the rule (\textit{Id}) applies.

The above reduction rules may create superpositions.  However,
such superpositions are not arbitrary.  Indeed, terms in a superposition
will only differ in positions containing the constants $0$ and $1$.
Otherwise they will have the same shape.  We may formalize this by defining
two terms to be \textit{congruent} if they coincide symbol by symbol
except possibly in positions containing $0$ or $1$.  It then follows
that
\begin{lem}
All terms in a superposition obtained via a reduction sequence from a
definite initial term are
congruent.
\label{congruent}
\end{lem}
\begin{proof}
The proof is
by a simple induction on the length of the reduction sequence, analyzing
the reduction rules case by case.
\end{proof}
Another case by case induction argument may be used to prove that reduction
preserves well-formedness.  More precisely
\begin{lem}
If $t$ is well-formed and $\ket{\His;\,t} \longto \sum_ic_i\,\ket{\His_i';\,t_i'}$,
then all terms $t_i'$ appearing in the resulting superposition are
well-formed.\footnote{Thanks to one of the referees for suggesting
improvements in the exposition of this section.}
\end{lem}
\noindent
Because terms appearing in a superposition have the same shape,
it makes sense to talk about specific subterms of the expression
in the computational register.
We can therefore formulate the following lemma:
\begin{lem}
Starting from a definite initial term,
any $!$-suspension subterm occurring during reduction
is definite with respect
to the computational basis.
\label{definite}
\end{lem}
\begin{proof}
This follows by induction on the length of the
reduction sequence.  The initial term is definite by assumption.
Assume that the lemma holds after $n$ steps.
We have argued that all terms in a superposition obtained from
a definite initial term are congruent.
They therefore have the
same structure of
subterms and the same reduction rule applies to them all.
Since we have argued that these terms are well-formed,
there are then
three ways in which we may obtain a $!$-suspension subterm after $n+1$ steps.
First, the suspension may not be part of the redex, in which case it is
included unmodified in the resulting expression.
Second, it may be the result
of beta reduction of an application of the form
$$
  (\lambda x.\,(\cdots x \cdots))\ (\cdots\, !t \cdots)
$$
where $!t$ is definite by the induction assumption.
The result is $(\cdots (\cdots\, !t \cdots) \cdots)$, where
$!t$ has been copied without modification.
Third, it may be the result
of beta reduction of an application of the form
$$
  (\lambda !x.\,\cdots\,!(\cdots x \cdots)\cdots)\ !t
$$
where $!t$ and $!(\cdots x \cdots)$ are definite by the induction
assumption.  This creates
a suspension $!(\cdots t \cdots)$, which is
definite because all its subterms are definite.  This completes the proof.
\end{proof}

It is worth pointing out that we cannot create possibly non-definite
suspensions by reducing terms like
$$
  (\lambda x.\,\cdots\,!(\cdots x \cdots)\cdots)\ (\Had\ 0)
$$
because $x$ is linear, which implies that $!(\cdots x \cdots)$ is not a
well-formed subterm.

\begin{lem}
Given a definite initial term,
the contents of the history track remains definite throughout reduction.
\end{lem}
\begin{proof}
We have argued that all terms in a superposition obtained from
a definite initial term are congruent.  They therefore have the
same structure of
subterms and the same reduction rule applies to them all.
Since our reduction rules allow only $!$-suspensions to be discarded
(which saves a copy in the history), and since $!$-suspensions are
always definite by the previous lemma, the result follows by induction
on the length of the reduction sequence.
\end{proof}

Since both the history and the shape of the term in the computational register
remain definite throughout reduction, we can state the following conclusion
\begin{cor}
Termination can be tested
without disturbing the computation by observing
the last term in the history.  When this term becomes equal to the placeholder $\ph$,
the result can be read off from the computational register.

Not only is termination a ``classical" property, but so is the entire
shape of the term itself, i.e., term shapes can be implemented in
classical memory, and only their data slots need to point to qubits on
a quantum device.\footnote{I would like to thank one of the referees
for suggesting this improved formulation.}
\end{cor}
The fact that the history remains definite
in $\lambda_q$ eliminates the
specific impediments to setting up an equational theory that were
pointed out in the previous section.  Indeed, since the state of the
computation is now always guaranteed to be a direct product $\ket{\His} \otimes \ket{c}$
of the history $\ket{\His}$ and the computational register $\ket{c}$, reduction
can never lead to a
computational register $\ket{c}$ that is in a mixed state.
In addition, since $\ket{\His}$ remains definite,
the restriction of the reduction rules to
the computational register will preserve the normalization.
We are therefore led to the following theorem:
\rulefig{
\brules
\begin{align*}
&\!\!\!\!\!\!\!\!\!\!\!\!\!\!\!\!\!{t_1 \longto t_1' \over
 (t_1\ t_2) \longto (t_1'\ t_2)} &&(\textit{app}_1)\\
&\!\!\!\!\!\!\!\!\!\!\!\!\!\!\!\!\!{t_2 \longto t_2' \over
 (v_1\ t_2) \longto (v_1\ t_2')} &&(\textit{app}_2)\\
(\lambda x.\, \Exp)\ v &\longto \Exp\,[v/x]
   &&(\beta) \\
(\lambda !x.\, \Exp)\ !t' &\longto \Exp\,[t'/x]
   &&(!\beta) \\
  \ket{c_U\ \phi} &\longto U\ket \phi
      &&(U)
\end{align*}
\erules
}{
Reduction rules for the quantum calculus $\lambda_q$  \label{qred}
}

\begin{thm}
In the quantum calculus $\lambda_q$, the evolution of the computational
register is governed by the reduction rules of figure \ref{qred}.
\end{thm}
\begin{proof}
This easily follows from a case-by-case analysis of the computational rules of
figure \ref{qoperational}.

For example, consider the rule ($!\beta_2$)
applied to a state of the form
$
\ket{\His}\otimes \ket{c}
$
where
$\ket{c}$ is a normalized superposition of the form
$$
\sum_i c_i \,\ket{(\cdots_i \,((\lambda !x.\, \Exp_i)\ !t')\,\cdots_i)}
$$
in which,
by lemma \ref{congruent}, all terms have the same structure and by lemma \ref{definite},
the subterm $!t'$ does not depend on $i$.  This then reduces to
\begin{align*}
&\ket{\His}\otimes \sum_i c_i\,\ket{(\ph\ph\ph((\lambda !x.\, \ph)\ !t')\ph\ph\ph);\,
(\cdots_i \, \Exp_i\,[t'/x] \,\cdots_i)} \\
&\qquad = \ket{\His;\,(\ph\ph\ph((\lambda !x.\, \ph)\ !t')\ph\ph\ph)}
\otimes \sum_i c_i\,
\ket{(\cdots_i  \,\Exp_i\,[t'/x] \,\cdots_i)}
\end{align*}
with the computational register in the normalized state
$\sum_i c_i\,
\ket{(\cdots_i\,  \Exp_i\,[t'/x]\, \cdots_i)}$, consistent with applying the reduction rule
$(\lambda !x.\, \Exp_i)\ !t' \longto \Exp_i\,[t'/x]$ to its contents.
\end{proof}

The big win is that we now have a simple set of reduction rules that can be used
to reason about the
computation without having to keep track of the history.

In order to define an equational theory for this calculus, we will simply define
a notion of equality that is compatible with the the reduction rules of figure
\ref{qred}.
Intuitively, reduction should be understood as a simple
algebraic operation of replacing subterms with
equal subterms.
However, we need to take into account that reduction may not take place inside
$!$-suspensions.

We therefore need to introduce algebraic rules
governing just when we can replace subterms in an expression with equal
subterms \cite{callbyvalue0, callbyvalue1}.
One way to do that is to introduce the notion of a \textit{term context},
which are expressions with a hole $[\:]$ in place of a subexpression:
\begin{align*}
  C,\, C_i&::= [\:]\ |\ (t\ C)\ |\ (C\ t)\ |\ (\lambda x.\, C)\
        |\ (\lambda !x.\, C)\ |\ \sum_i c_i\, C_i,
\end{align*}
where the last term denotes a superposition of shape-congruent contexts.
It is important to note that there are no contexts of the form $!(\cdots[\:]\cdots)$.  As a result,
subterms preceded by $!$ will be opaque in the sense that we will not be
able to perform substitutions under the $!$ sign.

\begin{defn}
\label{eqtheory}
The equational theory of $\lambda_q$ is the least equivalence relation $=$
containing the reduction relation ($\longto$) of figure \ref{qred}
and which is closed under
substitution in term contexts \cite{callbyvalue0, callbyvalue1}.
In other words
\begin{align*}
  {t_1 = t_2 \over C[t_1] = C[t_2]} &&(\textit{subst})
\end{align*}
where $C$ is an arbitrary term context and $C[t]$ denotes the textual
replacement of the hole in $C$ by the term $t$, extended by linearity
to superpositions of congruent terms and congruent contexts, i.e.,
for $t = \sum_i c_i\,t_i$, we define
$C[t] = \sum_i c_i\,C[t_i]$ and for $C = \sum_i c_i\,C_i$, we define
$C[t] = \sum_i c_i\,C_i[t]$.
\end{defn}

An alternative way of presenting the equational theory is by listing
a set of axioms and rules of inference as in figure \ref{eqproof}.
\rulefig{
\brules
\begin{align*}
 t = t & &&(\textit{refl})\\
 {t_1 = t_2 \over
 t_2 = t_1} & &&(\textit{sym}) \\
 {t_1 = t_2\quad t_2 = t_3 \over
 t_1 = t_3} & &&(\textit{trans}) \\
 {t_1 = t_2\quad t_3 = t_4 \over
 (t_1\ t_3) = (t_2\ t_4)} & &&(\textit{app}) \\
 {t_1 = t_2 \over
 \lambda x.\,t_1 = \lambda x.\,t_2} & &&(\textit{$\lambda_1$}) \\
 {t_1 = t_2 \over
 \lambda !x.\,t_1 = \lambda !x.\,t_2} & &&(\textit{$\lambda_2$}) \\
(\lambda x.\, \Exp)\ v = \Exp\,[v/x]
   & &&(\beta) \\
(\lambda !x.\, \Exp)\ !t' = \Exp\,[t'/x]
   & &&(!\beta) \\
\ket{(c_U\ \phi)} =  U\ket \phi
   &   &&(U)
\end{align*}
\erules
}{
Equational proof system for the quantum calculus $\lambda_q$  \label{eqproof}
}
These rules should again be understood as extending via linearity to
congruent superpositions of terms.
In this formulation, the rules (\textit{app}), ($\lambda_1$) and ($\lambda_2$) are together
equivalent to the term context substitution rule (\textit{subst}) above.
Again, there is no rule that permits substitutions inside $!$-suspensions.

\begin{thm}
In the quantum lambda calculus, the evolution of the
computational register proceeds by replacing terms by equal terms
according to the equational theory of definition \ref{eqtheory}.
\end{thm}
\begin{proof}
True by construction.
\end{proof}

\section{Recursion and a fixed point operator}

Recursive functions may be defined in the calculus $\lambda_q$
in a way analogous to that described in section \ref{classical}.
We simply replace $(t\ t)$ in equation (\ref{recurs1}) with $(t\ !t)$,
where now $t \equiv \lambda!f.\, (\,\cdots\, (f\ !f)\,\cdots)$.

Here we describe a related approach based on so-called fixed point
combinators. A fixed point operator suitable for the linear lambda calculus is
given by the following adaptation of the classical Turing combinator
\bcode
\fix \equiv
   (\!\begin{aligned}[t]
   &(\lambda !u.\,\lambda !f.\, (f\ !((u\ !u)\ !f))) \\
   !&(\lambda !u.\,\lambda !f.\, (f\ !((u\ !u)\ !f)))
   )
   \end{aligned}
\ecode
It is easy to check that under reduction
\begin{align*}
  \fix\ !t \longto t\ !(\fix\ !t)
\end{align*}
where the $!$-suspension prevents further reduction of the term in brackets.
Recursive functions can be defined as follows:  If
$$
  t \equiv \lambda !f.\,u
$$
then it easily follows that
$$
  \fix\ !t \longto u\,[(\fix\ !t)/f]
$$
In other words, $\fix\ !t$ copies itself into the body $u$ of $t$ under
reduction, as required for recursion.

\section{Examples of algorithms}
\label{sexamples}

We are now ready to formulate some algorithms in the quantum lambda calculus.
First, we reproduce some classical constructions, now decorated with the
proper nonlinearity annotations.

First, we introduce list constructors that will enable us to build
lists of linear values (qubits or structures containing qubits)
\begin{align*}
\empt &\equiv \lambda !x.\,\lambda !y.\,(x\ \id) \\
\cons &\equiv \lambda h.\,\lambda t.\,\lambda !x.\,\lambda !y.\,((y\ h)\ t)
\end{align*}
with abbreviations $h:t \equiv (\cons\ h\ t)$ and
$(x_1, \ldots, x_n) \equiv {x_1:x_2:\cdots :x_n:\empt}$ as before.
Since the arguments $!x$ and $!y$ above are nonlinear, we need to
redefine our $\case$ abbreviation as follows:
\bcode
\case \Exp_1 \of (\empt \to \Exp_2,\, h:t \to \Exp_3)
\ecode
now stands for
\bcode
\Exp_1\ !(\lambda !z.\,\Exp_2)\ !(\lambda h.\,\lambda t.\,\Exp_3)
\ecode
Deutsch's algorithm \cite{qtm1, nielsen} can be very simply expressed as
follows:
\bcode
    \deutsch\ U_f \longto
      \begin{aligned}[t]
        &\clet (x, y) = U_f\ ((\Had\ 0), (\Had\ 1)) \cin\\
        &\quad ((\Had\ x), y)
      \end{aligned}
\ecode
\fig{
\center{
\begin{picture}(93,30)
\put(8,10){\line(1,0){10}}
\put(28,10){\line(1,0){10}}
\put(8,25){\line(1,0){10}}
\put(28,25){\line(1,0){10}}
\put(0,9){\makebox{$\ket{1}$}}
\put(0,24){\makebox{$\ket{0}$}}
\put(18,5){\framebox(10,10){$H$}}
\put(18,20){\framebox(10,10){$H$}}
\put(38,5){\framebox(25,25){$U_f$}}
\put(63,10){\line(1,0){30}}
\put(63,25){\line(1,0){10}}
\put(83,25){\line(1,0){10}}
\put(73,20){\framebox(10,10){$H$}}
\end{picture}
}
}{
Deutsch's algorithm
}
Here the argument $U_f$ is assumed to be a function that takes
$(x, y)$ to $(x, y\oplus f(x))$, where $f$ is some (unknown)
function of one bit.  For example, if $f$ is the identity function,
then we should take $U_f$ to be $\cnot$. Indeed, the reader may check that
$$
  \ket{\deutsch\ \cnot} \longto \ket{1}\otimes\half\biggl(\ket0 - \ket1\biggr),
$$
where the first bit $1 = f(0)\oplus f(1)$ indicates that the function is balanced,
as required.

Let us write a simple expression that creates an EPR pair
\bcode
  \epr  \equiv \cnot\ ((\Had\ 0), 0)
\ecode
The quantum teleportation gate array with deferred measurement \cite{teleport, nielsen}
 can easily be translated into
the following code:
\fig{
\center{
\begin{picture}(118,45)
\put(0,9){\makebox{$\ket{0}$}}
\put(0,24){\makebox{$\ket{0}$}}
\put(0,39){\makebox{$x$}}
\put(8,10){\line(1,0){70}}
\put(8,25){\line(1,0){10}}
\put(18,20){\framebox(10,10){$H$}}
\put(28,25){\line(1,0){10}}
\put(38,9){\line(0,1){16}}
\put(38,10){\circle{3}}
\put(38,25){\circle*{3}}
\put(8,40){\line(1,0){30}}
\put(48,24){\line(0,1){16}}
\put(38,25){\line(1,0){20}}
\put(48,25){\circle{3}}
\put(48,40){\circle*{3}}
\put(38,40){\line(1,0){20}}
\put(58,35){\framebox(10,10){$H$}}
\put(68,40){\line(1,0){50}}
\put(48,25){\line(1,0){70}}
\put(78,5){\framebox(10,10){$X$}}
\put(83,15){\line(0,1){10}}
\put(98,5){\framebox(10,10){$Z$}}
\put(103,15){\line(0,1){25}}
\put(88,10){\line(1,0){10}}
\put(83,25){\circle*{3}}
\put(103,40){\circle*{3}}
\put(108,10){\line(1,0){10}}
\end{picture}
}
}{
Quantum teleportation
}
We create an EPR pair and
pass the first EPR qubit, along with the unknown qubit $x$ to be teleported,
to Alice.  The outcome $(x', y')$ of Alice's computation
then gets sent to Bob, who has access to the
second EPR qubit $e_2$
\bcode
  \teleport\ x \longto
    \begin{aligned}[t]
      &\clet (e_1, e_2) = \epr \cin \\
      &\clet (x', y') = \alice\ (x, e_1) \cin \\
      &\quad \bob\ (x', y', e_2)
    \end{aligned}
\ecode
Here
\bcode
   \alice\ (x, e_1) \longto
     \begin{aligned}[t]
       &\clet (x', y') = \cnot\ (x, e_1) \cin ((\Had\ x'), y')
     \end{aligned}
\ecode
and
\bcode
  \bob\ (x', y', e_2) \longto
    \begin{aligned}[t]
      &\clet (y'', e_2') = \cX\ (y', e_2) \cin \\
      &\clet (x'', e_2'') = \cZ\ (x', e_2') \cin \\
      &\quad (x'', y'', e_2'')
    \end{aligned}
\ecode
The outcome of the computation consists of the list of three qubits $(x'',
 y'', e_2'')$.
The teleported qubit is $e_2''$, but notice how linearity requires us
to keep the other two qubits in the answer.
The reader may check that throughout the computation, linear arguments are used exactly once.
Implementing the conditional operations $\cX$ and $\cZ$ in terms of the
primitive constants is left as an easy exercise.

Given recursion and lists, the $\map$ function, which applies
a given function $f$ to each element of a list, may be defined as
\bcode
\map\ !f\ \mathit{list} \longto
          \lcase {\mathit{list}}
                 {\empt}
                 {h:t} {(f\ h):(\map\ !f\ t)}
\ecode
The arguments $\mathit{list}$, $h$ and $t$ may refer to qubits
or data structures containing qubits and are therefore chosen linear.
The expression is well-formed because $\mathit{list}$, $h$ and $t$ are each used
exactly once.

It is now trivial to define a program that computes a
uniform superposition of a list of qubits by applying the
Hadamard gate to each qubit in the list:
\bcode
  \Had^{\otimes n}\ \mathit{list} \longto \map\ !\Had\ \mathit{list}
\ecode
For example, we may evaluate
$$
  \ket{\Had^{\otimes n}\ (0, 0)} \longto \half\bigg(\ket{(0, 0)} + \ket{(0, 1)}
     + \ket{(1, 0)} + \ket{(1, 1)}\biggr)
$$
Note that the well-formedness conditions may be somewhat subtle, as the
following example illustrates.  A naive attempt at defining an $\append$
function that concatenates two linear lists
\bcode
\append\ x\ y \longto
   \lcase {x}
          {y}
          {h:t} {h:(\append\ t\ y)}
\ecode
fails to be well-formed.  The problem can be seen by expanding the
$\case$ abbreviation
\bcode
x\ !(\lambda z.\,y)\ !(\lambda h.\,\lambda t.\,(h:(\append\ t\ y)))
\ecode
Since $y$ is a linear variable, we may not promote the $\lambda$ subterms
to nonlinear values with the prepended $!$.  An alternative definition that does work
is
\bcode
\append\ x\ y \longto
\left(
   \lcase {x}
          {(\lambda u.\, u)}
          {h:t} {\lambda u.\, (h:(\append\ t\ u))}
          \right)\ y
\ecode
Next we define a $\reverse$ function
\bcode
\reverse\ \textit{list} \longto
  \lcase {\textit{list}}
         {\empt}
         {h:t} {\append\ (\reverse\ t)\ (h)}
\ecode
The quantum Fourier transform \cite{qfourier1, qfourier2, shor}
can now be defined as a direct translation of
the corresponding quantum circuit \cite{nielsen} as follows:
\fig{
\center{
\begin{picture}(210,75)(-10,5)
  \put(-10,70){\line(1,0){10}}
  \put(0,65){\framebox(10,10){$H$}}
  \put(10,70){\line(1,0){10}}
  \put(20,65){\framebox(10,10){$R_2$}}
  \put(30,70){\line(1,0){10}}
  \put(40,65){\framebox(10,10){$R_3$}}
  \dottedline[$\cdot$]{2}(51,70)(59,70)
  \put(60,65){\framebox(10,10){$R_n$}}
  \put(70,70){\line(1,0){130}}

  \put(25,55){\line(0,1){10}}
  \put(25,55){\circle*{3}}
  \put(45,40){\line(0,1){25}}
  \put(45,40){\circle*{3}}
  \put(65,10){\line(0,1){55}}
  \put(65,10){\circle*{3}}

  \put(-10,55){\line(1,0){90}}
  \put(80,50){\framebox(10,10){$H$}}
  \put(90,55){\line(1,0){10}}
  \put(100,50){\framebox(10,10){$R_2$}}
  \dottedline[$\cdot$]{2}(111,55)(119,55)
  \put(120,50){\framebox(14,10){$R_{n-1}$}}
  \put(134,55){\line(1,0){66}}

  \put(105,40){\line(0,1){10}}
  \put(105,40){\circle*{3}}
  \put(127,10){\line(0,1){40}}
  \put(127,10){\circle*{3}}

  \put(-10,40){\line(1,0){150}}
  \put(140,35){\framebox(10,10){$H$}}
  \dottedline[$\cdot$]{2}(151,40)(159,40)
  \put(160,35){\framebox(14,10){$R_{n-2}$}}
  \put(174,40){\line(1,0){26}}

  \put(167,10){\line(0,1){25}}
  \put(167,10){\circle*{3}}

  \dottedline[$\cdot$]{2}(25,20)(25,30)

  \put(-10,10){\line(1,0){180}}
  \put(60,10){\line(1,0){110}}
  \dottedline[$\cdot$]{2}(171,10)(179,10)
  \put(180,5){\framebox(10,10){$H$}}
  \put(190,10){\line(1,0){10}}
\end{picture}
}}{
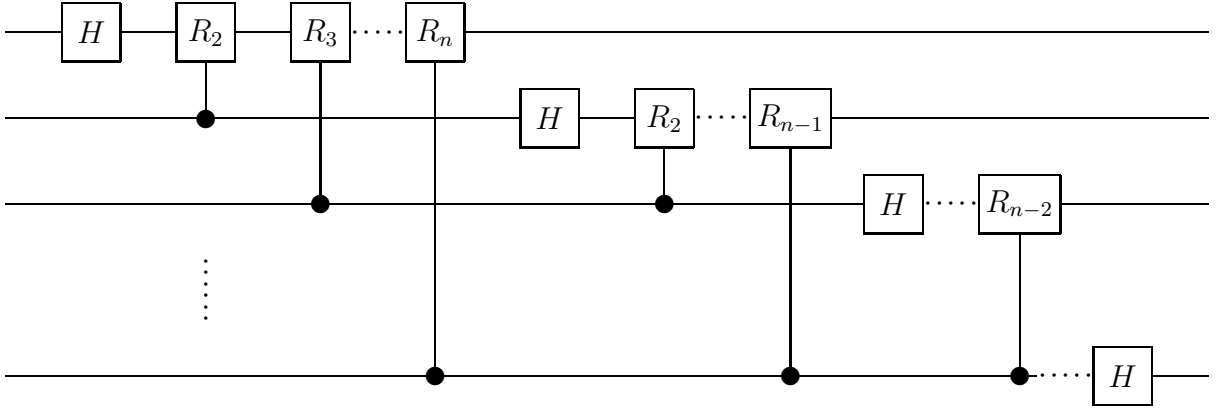
The quantum Fourier transform (without reversal)
}
\bcode
  \fourier\ \textit{list} \longto \reverse\ \fourier'\ \textit{list}
\ecode
where
\bcode
  \fourier'\ \textit{list} \longto
    \lcase {\textit{list}}
           {\empt}
           {h:t} {\begin{aligned}[t]
                  &\clet h':t' = \phases\ (\Had\ h)\ t\ !\un2\, \cin \\
                  &\quad h':(\fourier'\ t')
                  \end{aligned}}
\ecode
recursively applies the appropriate conditional phase operations to the
first qubit in the list, using the helper function
\bcode
\begin{aligned}
  &\phases\ \textit{target}\ \textit{controls}\ !n \\
  &\quad\longto
  \lcase {\textit{controls}}
         {(\textit{target})}
         {\textit{control}:t}
         {\begin {aligned}[t]
            &\clet
                (\textit{control}\,', \textit{target}\,') \\
                  &\qquad =  (\textit{\cphase}\ !n)\ (\textit{control}\,, \textit{target}\,) \cin
                  \\
            &\clet \textit{target}\,'':t' \\
            &\qquad = \phases\ \textit{target}\,'\ t\ !(\suc\  n)
               \cin \\
            &\quad \textit{target}\,'':\textit{control}\,':t'
          \end{aligned}}
\end{aligned}
\ecode
Here $(\textit{\cphase}\ !n)$ composes an appropriate combination of elementary gates
to implement
a conditional phase operation with phase $2\pi i / 2^n$.  Since this is essentially a
classical computation and depends on the particular set of primitive constants chosen,
we will not write it out here.

Note that we have assumed that the classical construction of the natural numbers
may be adapted to the quantum lambda calculus.  That this is possible for all
classical constructions follows from the fact that there
is an embedding of the
classical lambda calculus into the linear lambda calculus, as
shown in formula (\ref{translation}) of section \ref{universality}.

\section{Relating $\lambda_q$ to quantum Turing machines}
\label{universality}

In this section we will sketch a proof of the following theorem, leaving
a more rigorous analysis to future work:
\begin{thm}
The computational model
provided by the quantum lambda calculus
$\lambda_q$ is equivalent to the quantum Turing machine.
\end{thm}
\begin{proof}
First, we argue that the quantum lambda calculus $\lambda_q$ may be
efficiently simulated on a quantum Turing machine.

In $\lambda_q$ the current state of the computation consists of a superposition
of term sequences of the form $\His;\,t$, which may be
encoded as strings of symbols on the tape of the quantum Turing machine.
By lemma \ref{congruent},
term sequences in different
branches of the superposition are congruent, and
the same reduction rule will apply for all branches at each time
step.
The subset of $\lambda_q$ not involving quantum operations consists of a set of
reversible classical rewritings,
which can be unitarily and efficiently implemented on a quantum Turing
machine by \cite{qtm2, bv, bbbv}.  The fragment involving quantum operators
again involves simple classical rewritings followed by a unitary transformation
involving one or two symbols on the tape.  Once again, the methods of
\cite{qtm2, bv, bbbv} may be used to construct a quantum Turing machine that
can execute these transformations.  This completes the proof of the first
half of the equivalence.

Next we argue that a quantum Turing machine can be efficiently
simulated by the calculus $\lambda_q$.

Yao shows in \cite{yao} that for any quantum Turing machine $T$, there is
quantum circuit $C_{n,t}$ that efficiently simulates $T$ on inputs of size $n$
after $t$ steps.  The circuit family
$C_{n,t}$ may be efficiently constructed via a classical computation.
But $\lambda_q$ is universal for classical computation.  This
follows from the fact that the
classical call-by-value lambda calculus may be embedded in $\lambda_q$
via the following translation, adapted from \cite{linearcall}
\begin{equation}
\begin{aligned}
  (t_1\ t_2)^* &= ((\lambda !z.\, z)\ t_1^*)\ t_2^* \\
  x^* &= {!x} \\
  (\lambda x.\,t)^* &= {!(\lambda !x.\,t^*)}
\end{aligned}
  \label{translation}
\end{equation}
So, given the specification of a quantum Turing machine and an input of length $n$,
a classical computation in $\lambda_q$ first constructs a representation
of the appropriate quantum circuit family $C_{n,t}$.  It then follows the circuit
diagram and applies the appropriate quantum operations one by one to the input.
Since $\lambda_q$ has primitive quantum operations available corresponding to a
universal set of quantum gates, this proves the second half of the
equivalence.
\end{proof}

\section{Related work}

In a series of papers,
Henry Baker \cite{baker0, baker1, baker2} develops an untyped linear language
based on Lisp.  His language is
similar to the classical fragment of the lambda calculus developed in the
current article.  It served as the initial inspiration for the linear approach
followed here.

Ideas
stemming from linear logic have been used previously by Abramsky in the study of classical
reversible computation \cite{abramskyrev}.

One of the earlier attempts at formulating a language for quantum
computation was
Greg Baker's Qgol \cite{qgol}.  Its implementation (which remained incomplete)
used so-called uniqueness types (similar but not identical to our linear variables)
for quantum objects \cite{uniqueness}.
The language is not universal for quantum computation.

The language QCL, developed by \"Omer, is described in
\cite{omer1, omer2}.  QCL is an imperative language with
classical control structures combined with special operations on
quantum registers.  It provides facilities for inverting
quantum functions and for scratch space management.  No formal
program calculus is provided.
A simulator is publicly available.

Another imperative language, based on C++,
is the Q language developed by Bettelli, Calarco and Serafini \cite{bettelli}.
As in the case of QCL, no formal calculus is provided.
A simulator is also available.

A more theoretical approach is taken by Selinger in his description of
the functional language QPL \cite{selinger}.  This language has both
a graphical and a textual representation.  A formal semantics is provided.

The imperative language qGCL, developed by Sanders and Zuliani \cite{zuliani}, is
based on Dijkstra's guarded command language.  It has a formal semantics
and proof system.

A previous attempt to construct a lambda calculus for quantum computation is described
by Maymin in \cite{maymin1}.  However, his calculus appears to be
strictly stronger than the quantum Turing machine \cite{maymin2}.
It seems to go beyond quantum mechanics in that
it does not appear to have a unitary and reversible operational model,
instead relying on a more general class of transformations.
It is an open question whether the calculus is physically realizable.

A seminar by Wehr \cite{wehr} suggests that linear logic may be useful
in constructing a calculus for quantum computation within the mathematical
framework of Chu spaces.  However, the author stops short of developing
such a calculus.

Abramsky and Coecke describe a realization of a model of multiplicative linear
logic via the quantum processes of entangling and de-entangling by means of
typed projectors.  They briefly discuss how these processes can be represented
as terms of an affine lambda calculus \cite{abramskycoecke}.

\section{Conclusion}

In this article we developed a lambda calculus $\lambda_q$ suitable
for expressing and reasoning about quantum algorithms.
We discussed both its computational model and its
equational proof system.
We argued that the resulting calculus provides
a computational model equivalent to the quantum Turing machine
and is therefore universal for quantum computation.

There are many possible directions for future work.
The proof of Turing equivalence should be fleshed out.
Formal issues relating to consistency and semantics need to be
addressed further.
While our computational model
provides an operational semantics, the problem
of providing a denotational semantics is open.  The formalism of
\cite{kashefi} may be useful in this regard.

In this article, the introduction of a \textit{linear} calculus was motivated
by requiring consistency of its operational model with equational reasoning.
The fact that linear arguments, denoting quantum resources, may not be
duplicated suggests a separate motivation for linearity, not addressed here, based on
the no-cloning theorem \cite{noclone1, noclone2}.

While our calculus is untyped, it would be interesting to investigate
typed linear calculi with quantum primitives and,
via the Curry-Howard correspondence, the
corresponding generalizations of linear logic \cite{curry, howard}.
We might mention that there have been prior attempts to relate linear
logic to quantum mechanics, starting with a suggestion by Girard
\cite{girard, pratt, smets}.

On the practical side, the calculi described in this paper may be
used as a programming language for prototyping quantum algorithms.
Indeed,
the algorithms exhibited in this article were transcribed into Scheme
for testing.  The simulator, which was also written in Scheme,
is available upon request from the author.

It is our hope that the field of
quantum computation,
like its classical counterpart, may benefit from the insights provided
by the alternative computational model provided by
the quantum  lambda calculus.

\section*{Note added}

Since the first version of this paper was written, some progress has
been made by the author in devising a typed version, with accompanying
denotational semantics, of a fragment of the
quantum calculus described here \cite{myself}.

\section*{Acknowledgments}

    I would like to thank Prof.\ Antal Jevicki and the Brown
    University Physics department for their support.  I would also
    like to thank the two anonymous referees for their brilliant, detailed
    and thoughtful comments leading to various corrections and
    improvements in the exposition.


\begin{thebibliography}{99}
        \bibitem{qtm1}
            P. Benioff,
            \textit{The computer as a physical system:
            A microscopic quantum mechanical Hamiltonian model of
            computers as represented by Turing machines},
            \textit{J. Stat. Phys.} \textbf{22} (5) (1980), 563-591.
        \bibitem{qtm2}
            D. Deutsch,
            \textit{Quantum theory, the Church-Turing principle and
            the universal quantum computer},
            \textit{Proceedings of the Royal Society of London}
            \textbf{A 400} (1985), 97-117.
        \bibitem{circuits}
            D. Deutsch,
            \textit{Quantum Computational Networks},
            \textit{Proceedings of the Royal Society of London}
            \textbf{A 439} (1989),  553-558.
        \bibitem{yao}
            A. Yao,
            \textit{Quantum circuit complexity},
            in \textit{Proceedings of the 34th Annual Symposium on the
            Foundations of Computer Science},
            IEEE Computer Society Press, Los Alamitos, CA (1993) 352-361.
        \bibitem{nielsen}
            M.A. Nielsen and I.L. Chuang,
            \textit{Quantum Computation and Quantum Information},
            Cambridge University Press 2000.
         \bibitem{barendregt}
            H.P. Barendregt,
            \textit{The lambda Calculus},
            North Holland, revised edition (1984).
        \bibitem{mitchell}
            J.C. Mitchell, \textit{Foundations of programming languages},
            MIT press 1996.
        \bibitem{pierce}
            B.C. Pierce, \textit{Types and programming languages},
            MIT press 2002.
        \bibitem{davis}
            R.E. Davis, \textit{Truth, deduction and computation:
            logic and semantics for computer science},
            Computer Science Press, New York 1989.
        \bibitem{gunter}
            C.A. Gunter, \textit{Semantics of programming languages:
            structures and techniques},
             MIT press 1992.
        \bibitem{lisp}
            J. McCarthy, \textit{Recursive functions of symbolic expressions
            and their computation by machine, part I},
            \textit{Communications of the ACM} \textbf{3} (4) (1960), 184-195.
        \bibitem{epl}
            D.P. Friedman, M. Wand, C.T. Haynes,
            \textit{Essentials of programming languages},
            MIT press 1992.
        \bibitem{ml}
            L.C. Paulson, \textit{ML for the working programmer},
            Cambridge University Press, 1996.
        \bibitem{haskell}
            P. Hudak, \textit{The Haskell school of expression},
            Cambridge University Press 2000.
        \bibitem{backus}
            J. Backus, \textit{Can programming be liberated from the
            von Neumann style?  A functional style and its algebra of
            programs},
            \textit{Communications of the ACM} \textbf{21} (8), 1978.
        \bibitem{myself}
            A. van Tonder,
            \textit{Quantum computation, categorical semantics and
            linear logic},
            ArXiv.org e-print archive: arXiv:quant-ph/0312174 (2003).
        \bibitem{church1}
            A. Church, \textit{An unsolvable problem in elementary number theory},
            \textit{American Journal of Mathematics} \textbf{58} (1936) 354-363.
        \bibitem{church2}
            A. Church, \textit{The calculi of lambda conversion},
            Princeton University Press 1941.
        \bibitem{turing}
            A.M. Turing,
            \textit{On computable numbers, with an application to
            the Entscheidungsproblem},
            \textit{Proc. London Math. Soc.} (2) \textbf{42} (1936) 230-265;
            Corrections in
            \textit{Proc. London Math. Soc.} (2) \textbf{43} (1937) 544-546.
        \bibitem{callbyvalue0}
            G. Plotkin,
            \textit{Call-by-name, call-by-value and the $\lambda$-calculus},
            \textit{Theoretical Computer Science} \textbf{1} (1) (1976), 125–159.
        \bibitem{callbyvalue1}
            M. Felleisen and R. Hieb,
            \textit{A revised report on the syntactic theories of
            sequential control and state},
            \textit{Theoretical Computer Science} \textbf{103} (2) (1992) 235-271.
        \bibitem{oleg}
            O. Kiselyov,
            \textit{Many faces of the fixed-point combinator},
            online article
            http://okmij.org/ftp/Computation/fixed-point-combinators.html,
            (Oct 1999).
        \bibitem{bennettrev}
            C.H. Bennett,
            \textit{Logical reversibility of computation},
            \textit{IBM J. Res. Develop.} \textbf{17} (1973), 525–532.
        \bibitem{gates1}
            A. Barenco, C.H., Bennett, R. Cleve,
            D.P. DiVincenzo, N. Margolus, P. Schor,
            T. Sleator, J. Smolin and H. Weinfurter,
            \textit{Elementary gates for quantum computation},
            \textit{Phys. Rev.} \textbf{A52} (1995), 3457-3467.
        \bibitem{gates2}
            P.O. Boykin, T. Mor, M. Pulver, V. Roychowdhury and F. Vatan,
            \textit{On universal and fault-tolerant quantum computing},
            arXiv:quant-ph/9906054 (1999).
        \bibitem{abramsky}
            S. Abramsky,
            \textit{Computational interpretations of linear logic},
            \textit{Theoretical Computer Science}
            \textbf{111} (1-2) (1993) 3-57.
        \bibitem{linearsyn}
            P. Wadler,
            \textit{A syntax for linear logic},
            in \textit{Mathematical Foundations of Programming Semantics:
             9th International Conference, New Orleans, LA, Proceedings}
             \textbf{802} Springer-Verlag, New York (1993) 513-529.
        \bibitem{linearcall}
           J. Maraist, M. Odersky, D. Turner and P. Wadler,
           \textit{Call-by-name, call-by-value, call-by-need,
           and the linear lambda calculus},
           in
           \textit{11th International
            Conference on the Mathematical Foundations of
            Programming Semantics}, New
            Orleans, Louisiana, March-April 1995.
        \bibitem{seely}
            R.A.G. Seely,
            \textit{Linear logic, *-autonomous categories, and cofree
            coalgebras},
            in \textit{Categories in Computer Science and Logic},
            June 1989,
            AMS Contemporary Mathematics 92.

        \bibitem{girard}
            J.-Y. Girard,
            \textit{Linear logic},
            \textit{Theoretical Computer Science} \textbf{50} (1987) 1-102.
        \bibitem{lineartaste}
            P. Wadler,
            \textit{A taste of linear logic},
            in \textit{Proceedings of the 18th International Symposium
            on Mathematical Foundations of Computer Science, Gd{\'a}nsk},
            Springer-Verlag, New York (1993).

        \bibitem{TurnerWadlerOperational}
            D.N. Turner and P. Wadler,
            \textit{Operational interpretations of linear logic},
            \textit{Theoretical Computer Science}
            \textbf{227} (1-2) (1999) 231-248.
        \bibitem{chirimar}
            J. Chirimar, C.A. Gunter and J.G. Riecke,
            \textit{Reference counting as a computational
               interpretation of linear logic},
            \textit{Journal of Functional Programming} \textbf{6} (2) (1996) 195-244.
        \bibitem{teleport}
            C.H. Bennett, G. Brassard, C. Cr\'epeau, R. Jozsa, A. Peres,
            and W. Wootters,
            \textit{Teleporting an unknown quantum state via dual classical and EPR channels},
            \textit{Phys. Rev. Lett.} \textbf{70} (1993) 1895-1899.
        \bibitem{qfourier1}
            D. Coppersmith, \textit{An approximate Fourier transform useful
            in quantum factoring},
            \textit{IBM Research Report} RC 19642 (1994).
        \bibitem{qfourier2}
            A. Ekert and R. Jozsa,
            \textit{Shor's quantum algorithm for factorizing numbers},
            \textit{Rev. Mod. Phys.} \textbf{68} (1996), 733-753.
        \bibitem{shor}
            P.W Shor.,
            \textit{Algorithms for quantum computation: discrete log and factoring},
            in \textit{Proceedings of the 35th IEEE FOCS},
            (1994) 124–134; \newline
            P.W. Shor,
            \textit{Polynomial-time algorithms for prime factorization
            and discrete logarithms on a quantum computer},
            \textit{SIAM Journal on Computing} \textbf{26} (5) (1997),
            1484-1509.
        \bibitem{bv}
            E. Bernstein and U. Vazirani, \textit{Quantum complexity theory},
            \textit{SIAM J. Computing} \textbf{26} (1997), 1411-1473.
        \bibitem{bbbv}
            C.H. Bennett, E. Bernstein, G. Brassard, U. Vazirani,
            \textit{Strengths and weaknesses of quantum computing},
            \textit{SIAM Journal on Computing} \textbf{26} (5) (1997) 1510-1523.

        \bibitem{baker0}
            H.G. Baker,
            \textit{Lively linear Lisp -- 'Look Ma, no garbage!'},
            \textit{ACM Sigplan Notices} \textbf{27} (8) (1992), 89-98.
        \bibitem{baker1}
            H.G. Baker,
            \textit{A 'Linear Logic' quicksort},
            \textit{ACM Sigplan Notices} \textbf{29} (2) (1994), 13-18.
        \bibitem{baker2}
             H.G. Baker,
             \textit{'Use-once' variables and linear objects --
             storage management, reflection and multi-threading},
             \textit{ACM Sigplan Notices} \textbf{30}, 1 (1995), 45-52.
        \bibitem{abramskyrev}
            S. Abramsky,
            \textit{A structural approach to reversible computation},
            Programming Research Group Research Report RR-01-09, Oxford
            University (2001).
        \bibitem{qgol}
            G.D. Baker,
            \textit{``Qgol'': A system for simulating quantum computations:
            theory, implementation and insights},
            Honours thesis, Macquarie University (1996).
        \bibitem{uniqueness}
            E. Barendsen and S. Smetsers,
            \textit{Conventional and uniqueness typing in graph rewrite systems},
             Computing Science Institute, University of Nijmegen,
             Technical report CSI-R9328 (December 1993).
        \bibitem{omer1}
            B. \"{O}mer,
            \textit{A procedural formalism for quantum computing},
            Master thesis, Technical University Vienna (1998).\newline
            http://tph.tuwien.ac.at/~oemer/qcl.html
        \bibitem{omer2}
            B. \"{O}mer,
            \textit{Classical concepts in quantum programming},
            quant-ph/0211100 (2002).
        \bibitem{bettelli}
            S. Bettelli, T. Calarco and L. Serafini,
            \textit{Towards and architecture for quantum programming},
            \textit{Eur. Phys. J.} \textbf{D 25} (2) (2003), 181-200.
            arXiv.cs.PL/0103009 (2001).
        \bibitem{selinger}
            P. Selinger,
            \textit{Towards a quantum programming language},
            to appear in \textit{Mathematical Structures in Computer Science}
            (2003) 45 pages.
        \bibitem{zuliani}
            J.W. Sanders and P. Zuliani,
            \textit{Quantum programming}
            in \textit{Mathematics of Program Construction}, Springer LNCS,
            1837:80-99, (2000).
        \bibitem{maymin1}
            P. Maymin,
            \textit{Extending the lambda calculus to express randomized and
            quantumized algorithms},
            quant-ph/9612052 (1996).
        \bibitem{maymin2}
            P. Maymin,
            \textit{The lambda-q calculus can efficiently simulate quantum computers},
            quant-ph/9702057 (1997).
        \bibitem{wehr}
            M. Wehr,
            \textit{Quantum computing: A new paradigm and its type theory},
            Talk held at
            \textit{Quantum Computing Seminar, Lehrstuhl Prof. Beth,
                  Universit\"at Karlsruhe} (1996).

        \bibitem{abramskycoecke}
            S. Abramsky and B. Coecke,
            \textit{Physical traces: Quantum vs. classical information processing},
            \textit{Electronic Notes in Theoretical Computer Science}
            \textbf{69} (2003),
            arXiv:cs.CG/0207057 (2002).
        \bibitem{kashefi}
            E. Kashefi,
            \textit{Quantum domain theory - definitions and applications},
            arXiv:quant-ph/0306077 (2003).
        \bibitem{noclone1}
            D. Dieks,
            \textit{Communication by EPR devices},
            \textit{Phys. Lett. A} \textbf{92} (1982), 271-272.
        \bibitem{noclone2}
            W.K. Wootters and W.H. Zurek,
            \textit{A single quantum cannot be cloned},
            \textit{Nature} \textbf{299} (1982), 802-803.
        \bibitem{curry}
            H.B. Curry and R. Feys,
            \textit{Combinatory Logic},
            Volume 1,
            North Holland, 1985.  Second edition, 1968.
        \bibitem{howard}
            W.A. Howard,
            \textit{The formulas-as-types notion of construction},
            in J.P. Seldin and J.R. Hindley, editors,
            \textit{To H.B. Curry: Essays on Combinatory Logic,
            Lambda Calculus, and Formalism}, Academic Press, New York 1980,
            pp. 479-490.  Reprint of 1969 article.
        \bibitem{pratt}
            V. Pratt,
            \textit{Linear logic for generalized quantum mechanics},
            in \textit{Proc. of Workshop on Physics and Computation (PhysComp'92)},
            IEEE, Dallas (1992), 166-180.
        \bibitem{smets}
             S. Smets,
             \textit{What has Operational Quantum Logic to do with Linear Logic?},
             presented at the Logic and Interaction Week 3,
             Marseilles, France, February 2002.

    \end{thebibliography}
\end{document}